**A generalized-growth model to characterize the early ascending phase of infectious disease outbreaks**

Cécile Viboud[1], Lone Simonsen[2,3], Gerardo Chowell[1,4]

[1] Division of International Epidemiology and Population Studies, Fogarty International Center, National Institutes of Health, Bethesda, MD, USA

[2] Department of Public health, University of Copenhagen, Copenhagen, Denmark

[3] Department of Global Health, School of Public Health and Health Services, George Washington University, Washington DC, USA

[4] School of Public Health, Georgia State University, Atlanta, GA, USA

**Corresponding author:**

Gerardo Chowell, PhD
School of Public Health, Georgia State University, Atlanta, GA, USA
Division of International Epidemiology and Population Studies, Fogarty International Center, National Institutes of Health, Bethesda, MD, USA
Email: gchowell@gsu.edu

Body word count: 3721

Abstract word count: 298

**Abstract**

**Background:**
A better characterization of the early growth dynamics of an epidemic is needed to dissect the important drivers of disease transmission, refine existing transmission models, and improve disease forecasts.

**Materials and Methods:**
We introduce a 2-parameter generalized-growth model to characterize the ascending phase of an outbreak and capture epidemic profiles ranging from sub-exponential to exponential growth. We test the model against empirical outbreak data representing a variety of viral pathogens in historic and contemporary populations, and provide simulations highlighting the importance of sub-exponential growth for forecasting purposes.

**Results:**
We applied the generalized-growth model to 20 infectious disease outbreaks representing a range of transmission routes. We uncovered epidemic profiles ranging from very slow growth ($p=0.14$ for the Ebola outbreak in Bomi, Liberia (2014)) to near exponential ($p>0.9$ for the smallpox outbreak in Khulna (1972), and the 1918 pandemic influenza in San Francisco). The foot-and-mouth disease outbreak in Uruguay displayed a profile of slower growth while the growth pattern of the HIV/AIDS epidemic in Japan was approximately linear. The West African Ebola epidemic provided a unique opportunity to explore how growth profiles vary by geography; analysis of the largest district-level outbreaks revealed substantial growth variations (mean $p=0.59$, range: 0.14-0.97). The districts of Margibi in Liberia and Bombali and Bo in Sierra Leone had near-exponential growth, while the districts of Bomi in Liberia and Kenema in Sierra Leone displayed near constant incidences.

**Conclusions:**
Our findings reveal significant variation in epidemic growth patterns across different infectious disease outbreaks and highlights that sub-exponential growth is a common phenomenon, especially for pathogens that are not airborne. Sub-exponential growth profiles may result from heterogeneity in contact structures or risk groups, reactive behavior changes, or the early onset of interventions strategies, and consideration of "deceleration parameters" may be useful to refine existing mathematical transmission models and improve disease forecasts.

## Introduction

Identifying signature features of the growth kinetics of an outbreak can be useful to design reliable models of disease spread and understand important details of the transmission dynamics of an infectious disease [1]. However, ideal data are typically not available; rather there will be an absence of high-resolution epidemiological datasets needed to characterize transmission pathways in key settings, e.g., transmission trees in hospitals, schools, households [2, 3]). The force of infection in mathematical transmission models is typically estimated using time-series data that describe epidemic growth as a function of time (e.g., [4-9]). In fact, during the 2003 SARS (Severe Acute Respiratory Syndrome) threat, the 2009 A/H1N1 influenza pandemic, and the 2013-15 Ebola epidemic in West Africa, aggregated case series at the national or district level were the primary sources of data available for model calibration (e.g. [10-13]). A better understanding of observed epidemic growth patterns across different pathogens and across temporal and social contexts could prove useful to improve our ability to design disease transmission models, including the important task of forecasting the likely final size of the epidemic (morbidity, mortality impact), as well as to assess the effects of control interventions.

Classical compartmental transmission models assume exponential growth during the early phase of a well-mixed epidemic[14]. In a recent article [1] we reported that the initial apparently exponential spread of the 2013-15 Ebola epidemic in West Africa was in fact a composition of local asynchronous outbreaks at the district or county level that each displayed sub-exponential growth patterns during at least 3 consecutive disease generations [1]. In semi-logarithmic scale, exponential growth is evident if a straight line fits well several consecutive disease generations of the epidemic curve, whereas a strong downward curvature in semi-logarithmic scale is indicative of sub-exponential growth. Here, we introduce a generalized model with an “deceleration” parameter that modulates growth and helps quantify departure from exponential theory, allowing for behaviors ranging from constant to exponentially-growing incidences [15]. Our simple quantitative framework is useful for public health decision-making as it provides a time-varying assessment of growth rates and informs the likely “signature feature” of the threat as well as the type and intensity of interventions required for effective mitigation. We illustrate our method using a diverse set of historic and contemporary outbreaks of acute viral and bacterial pathogens, including the recent West African Ebola virus epidemic, focusing on local outbreaks. Our results underscore the high sensitivity of epidemic size to small variations in the "deceleration" parameter.

## Materials and Methods

### Data sources

We characterized the initial epidemic growth patterns in various infectious disease incidence time series including pandemic influenza, measles, smallpox, bubonic plague, foot-and-mouth disease (FMD), HIV/AIDS, and Ebola (Table 1). The temporal resolution of the datasets varied from daily, weekly, to annual. These selected outbreak data represent a convenience sample encompassing a range of pathogens, geographic contexts, and time periods. For each outbreak, the onset week corresponds to the first observation associated with a monotonic increase in incident cases, up to the peak incidence.

### Generalized epidemic growth model

The growth pattern of infectious disease outbreaks has been extensively studied using models that assume exponential growth dynamics in the absence of control interventions (e.g, classical compartmental models [14, 16]). Hence, the cumulative number of cases, C(t), grows according to the equation: $C(t) = C(0)e^{rt}$ where r is the growth rate per unit of time, t denotes time, and C(0) is the number of cases at the start of the outbreak. Here the growth rate "r" is related to $R_0$ as derived from classic SIR-type compartmental transmission models, e.g., for the simple SIR (susceptible-infected-removed) model, $R_0 = 1 + r/\gamma$ where $1/\gamma$ is the mean infectious period [14]. However, slower-than-exponential epidemic growth is expected in settings that involve highly constrained population contact structures with infectious diseases that spread via close contacts (e.g., sexually-transmitted infectious diseases, smallpox, Ebola)[1]. To relax the assumption of exponential growth, we use a simple generalized model [15] from the field of demography (e.g., [17]), following:

$$\frac{dC(t)}{dt} = C'(t) = rC(t)^p$$

where C'(t) describes the incidence curve over time t, the solution C(t) describes the cumulative number of cases at time t, r is a positive parameter denoting the growth rate (1/time), and p$\in [0,1]$ is a "deceleration of growth" parameter. If p=0, this equation describes constant incidence over time and the cumulative number of cases grows linearly while p=1 models exponential growth dynamics (i.e., Malthus equation). Intermediate values of p between 0 and 1 describe sub-exponential (e.g. polynomial) growth patterns. For example, if p=1/2 incidence grows linearly while the cumulative number of cases follows a quadratic polynomial. If p=2/3 incidence grows quadratically while the cumulative number of cases fits a cubic polynomial. For sub-exponential growth (i.e., 0<p<1) the solution of this equation is given by the following polynomial of degree *m* [15]:

$$C(t) = \left(\frac{r}{m}t + A\right)^m$$

where *m* is a positive integer, and the "deceleration of growth" parameter is given by p=1-1/m. [15] *A* is a constant that depends on the initial condition, C(0). Specifically, $A = \sqrt[m]{C(0)}$. Furthermore, for sub-exponential growth dynamics, the relative growth rate, $[dC(t)/dt]/C(t) \propto m/t$, decreases inversely with time while the doubling time $T_d \propto t(\ln 2)/m$ increases proportionally with time (Figure 1)[1]. This differs from the constant doubling time that characterizes exponential growth. Here, we do not consider faster than exponential growth (i.e., super-exponential growth), for which p exceeds 1.0 [15].

**Parameter estimation**

Parameters r and p can be jointly estimated through nonlinear least-square curve fitting to the case incidence curve modeled by equation C'(t), in the first few generations of disease spread. For this purpose, we used the Levenberg-Marquardt algorithm implemented in MATLAB (The Mathworks, Inc.) as in prior studies (e.g, [18]). The initial number of cases C(0) was fixed according to the first observation. We estimate r and p during the initial epidemic growth phase comprising approximately 3-5 generations of disease transmission when the proportion of susceptible individuals in the population approximates its initial value. The mean generation interval has been estimated at ~3-5 days for influenza [19], about two weeks for measles [20], smallpox [13], and Ebola [21], about one week for pneumonic plague [22], ~3-7 days for foot-and-mouth disease[23] and has been estimated at ~4 years for HIV/AIDS [24].

**Confidence intervals**

Confidence intervals for the model parameter estimates were constructed by simulating 200 realizations of the best-fit curve C'(t) using parametric bootstrap with a Poisson error structure, as in prior studies [18, 25]. Parameters r and p were then estimated from each of 200 simulated epidemic curves to derive nominal 95% confidence intervals.

**Simulation study**

Our analysis of estimating parameters "r" and "p" for the generalized-growth model from simulated data indicate that the parameter uncertainty as measured by the width of the 95% confidence interval of the parameter estimates is reduced by larger case numbers, so that it typically remains higher the smaller the "deceleration of growth" parameter p. Further, parameter uncertainty is reduced as more cases are reported over time during the initial epidemic growth phase (Supplementary Figure S1).

If the case series are the only information available on the temporal progression of the epidemic then the most general choice to model the variability in case series is the Poisson distribution where the mean equals the variance, and which we employ in our analyses here. If additional information on the magnitude of fluctuations in case incidence over time were known, then a generalized distribution could be used such as the negative binomial distribution, which allows for greater variance and better accounts for over-dispersed data.

**Model comparison**

We statistically assessed the improvement in goodness of fit provided by the generalized growth model with two parameters over the simpler exponential growth model with a single parameter (i.e., p=1) through an F test [26]. We used a confidence level at 0.1.

For each infectious disease outbreak dataset (Table 1), we generated estimates and their corresponding uncertainty bounds for parameters r and p as the length of the initial epidemic growth phase varies. We displayed uncertainty in estimates of r and p as a function of the epidemic growth phase length, assessed the goodness of fit provided by the generalized-growth model as compared to the simpler exponential growth model where p=1, and showed representative fits of the generalized-growth model to incidence data.

## Results

Simulations indicate that the generalized-growth model supports different epidemic growth profiles, as the "deceleration" parameter (p) is varied between zero and one (Figure 1). These profiles include linear incidence (i.e., p=0.5), concave-up incidence (p>0.5), and concave-down incidence (p<0.5) patterns whereas the relative growth rates decline inversely with time (Figure 1). Moreover, epidemic size is predicted to be highly sensitivity to small variations in the deceleration parameter p, as shown in Figure 2.

Overall, our analysis of empirical disease data using the generalized-growth model revealed a diversity of profiles across infectious disease outbreaks, with most estimates of p displaying substantial uncertainty (overall mean p=0.61 S.D=0.26) (Table 1, Figure 3). Figures 4-9 display a representative set of model fits to the 19 infectious disease outbreaks.

Estimates of p ranged from 0.14 for the Ebola outbreak in Bomi, Liberia (2014), reflecting a slow epidemic growth pattern, to 0.85 for a major plague epidemic in Bombay (1905), 0.95 for the smallpox outbreak in Khulna (1972), and 0.98 for pandemic influenza in San Francisco (1918), consistent with near-exponential growth dynamics (Figure 4-5).

The foot-and-mouth disease outbreak in Uruguay at the farm level displayed a low mean estimate of p at 0.4, characterizing a profile of slower growth (Figure 6 and Figure 1). We found evidence for sub-exponential growth in annual incidence data on the HIV/AIDS epidemic in Japan (1985-2012) and New York City (1982-2002), a disease that is largely spread through sexual contact via bodily fluids. The mean estimate of the deceleration parameter p for time series of new HIV/AIDS cases in Japan was estimated at about 0.5 which is consistent with an approximately linear growth pattern (Figure 7) whereas the AIDS case reports in New York City showed slightly faster growth pattern with mean p estimated at 0.57.

The wealth of Ebola data available at the district level for the 2013-15 epidemic in West Africa provides a good opportunity to gauge variations in the growth profile of a given pathogen over space. Estimates of p across local Ebola outbreaks varied substantially (mean=0.59, range: 0.14-0.97). The districts of Margibi in Liberia, and Bombali and Bo in Sierra Leone, displayed near exponential growth (p close to 1), while the districts of Bomi in Liberia and Kenema in Sierra Leone displayed particularly slow growth (p near 0.1) (Figure 3). Of particular interest is the shape of the epidemic in the district where the 2013-15 Ebola epidemic is assumed to have originated, in Gueckedou, Guinea. We estimate an intermediate pattern of growth there, albeit with wide uncertainty (p=0.71 95% CI: 0.31, 1.0). Further, Montserrado, the Liberian capital, displayed moderate values of p (0.60 95% CI: 0.46, 0.77) (Figure 8), which is in close agreement with those of Western Area Urban and Western Area Rural in Sierra Leone with mean p at 0.46-0.68 (Figures 3 and 9).

## Discussion

To improve the ability of epidemiological models to capture the trajectory and impact of pandemics and epidemics, we have introduced a generalized-growth model to characterize the epidemic ascending phase, allowing to capture a range of growth profiles ranging from sub-exponential (e.g., polynomial) to exponential. Our approach provides a simple 2-parameter quantitative framework to assess epidemic growth patterns over time. This framework could also be useful to assess shifts in epidemic growth patterns resulting from mitigation during an outbreak, as a consequence of population behavior changes or public health control interventions that affect transmission (e.g., increased bed capacity, school closings). We illustrate the approach through simulations and applications to various empirical infectious disease datasets representing directly and sexually transmitted viral pathogens, including Ebola, pandemic influenza, smallpox, plague, measles, foot-and-mouth disease, and HIV/AIDS. Our findings indicate significant variability in epidemic growth patterns across pathogens and geographic settings (Table 1) and demonstrate high sensitivity of epidemic size estimates to small variations in the "deceleration" parameter of our generalized-growth model.

We have previously hypothesized that the sub-exponential epidemic growth patterns could arise from a combination of factors [1] that include: 1) a spatially constrained contact structure (e.g., high clustering levels)[27-31] directly linked to the epidemiological characteristics of the disease including the direct-contact (i.e., non-airborne) transmission mode of diseases such as Ebola and HIV/AIDS and the fact that individual infectiousness and disease severity are frequently correlated and 2) the role of population behavior changes and control interventions particularly during the later epidemic stages [28, 32, 33]. Other factors shaping sub-exponential growth patterns could include substantial heterogeneity in susceptibility and infectivity in the underlying population that further distorts the contact network structure relevant to disease spread. Faithfully capturing the baseline transmission pattern of an unfolding infectious disease outbreak is needed to prospectively quantify the effect of control measures. In particular, different epidemic growth patterns could support different intervention strategies.

Perhaps it is not surprising that the daily growth phase of the 1918 influenza pandemic in San Francisco and the weekly growth phase of the bubonic plague epidemic in Bombay in 1905 are close to exponential during several early generations of the disease, consistent with reported excellent fits of SEIR-type models reported in [18] and [34], respectively. It would be interesting to conduct a systematic analysis of growth patterns of pandemic and seasonal influenza epidemics in particular areas of the world with reliable and space- and time-resolved incidence datasets. We hypothesize that estimates of p would be lower for seasonal influenza epidemics and/or that rapid shifts in this parameter occur during seasonal influenza epidemics, reflecting the combined and complex benefits of underlying immunity, school activity patterns, and other behavioral changes in response to influenza epidemics [35].

We found evidence for sub-exponential growth in annual incidence data on the HIV/AIDS epidemic in Japan (1985-2012) and New York City (1982-2002), a disease that is largely spread through sexual contact via bodily fluids, quite different from the more democratic and airborne spread of pandemic influenza. These growth patterns are in line with an earlier study that characterized the cumulative number of AIDS cases in the United States via a cubic-polynomial growth function, a pattern that was explained by risk behavior-based transmission dynamics [36-38].

Our findings for the smallpox outbreak in Khulna municipality, Bangladesh (1972) point to near-exponential growth dynamics during the early epidemic phase of this epidemic. It is likely that this particular epidemic growth pattern is representative of the upper bound of expected growth patterns of smallpox, a disease that spreads mostly via close contact which facilitates its control and the implementation of ring vaccination strategies [39]. Yet, further analysis of historical smallpox datasets is warranted to better understand the dynamics of this infectious disease of the past. A limitation is that epidemiologic data have been collected on a coarse temporal resolution (e.g. yearly or monthly) which complicates the analysis of epidemic growth for an acute infection like smallpox [40].

Our analysis of subnational data of the 2013-15 Ebola epidemic in West Africa revealed wide variation in epidemic growth patterns ranging from slow to near-exponential growth dynamics (Figure 3). A previous visualization of the West African Ebola epidemic growth patterns at the district or county level revealed rapid epidemic saturation within just a few generation intervals of the disease [1]. While these results should be important to guide the design of mathematical models of Ebola transmission dynamics and control, a better understanding of the factors that shaped Ebola epidemic growth patterns including spontaneous behavior changes and increased healthcare infrastructure[41] are needed to guide the development of mechanistic transmission models that would allow the evaluation of specific control policies. Since the deceleration parameter is important to characterize the early epidemic growth phase, one could expect our generalized-growth model to fare better than classical SEIR models for short-term forecasting of epidemic growth.

We have analyzed here a single epidemic curve of measles in London in early 1948, which indicates a moderate growth pattern (p=0.5). A systematic analysis of epidemic growth patterns over multiple epidemic years (e.g, during the pre-vaccination area in England and Wales) is warranted in order to better understand the historical variation in growth patterns generated by measles epidemics [42]. More broadly, it would be interesting to compare epidemic growth patterns of childhood infectious diseases particularly in regions with solid epidemiological surveillance.

The moderate growth pattern of the 2001 FMD epidemic in Uruguay is consistent with slower spreading disease probably as a consequence of spatial movement, previous immunity in the population and possibly the impact of control interventions during the early epidemic phase [26]. It would be interesting to compare this epidemic growth pattern with that of the well-documented 2001 FMD epidemic in the United Kingdom [43, 44].

It is common practice to rely on epidemic models largely based on homogenous mixing assumptions, whereby the initial growth phase follows an exponential growth during several generations of the disease (e.g. [14, 18, 45, 46]). Estimates of the basic reproduction number $R_{0,}$ a key quantity for disease control, can be obtained by fitting these models to time-series of case reports during the early epidemic ascending phase [45] [46, 47]. Our findings indicate the assumption of exponential growth may be inappropriate, and sub-exponential growth patterns should be considered in parameterization of mathematical transmission models. In fact, Liu et al. (1987) [48] incorporated phenomenological parameters in the bilinear incidence rates of the mechanistic SIR model, as a way to expand the range of plausible dynamical epidemic behaviors. Similarly, Grenfell and Finkenstädt (2000)[49] developed the well-known TSIR modeling framework that includes a phenomenological parameter to capture non-homogeneous mixing in the analysis of infectious disease time-series data [42, 49]. A necessary condition for validating a transmission model is that the model is able to reproduce growth patterns that are consistent with observed epidemiological outbreak data [50]; and hence, knowledge about the extent of exponential growth in a given epidemic is useful to guide the design of transmission models. Further, model refinements relating to sub-exponential growth dynamics may be particularly important if these models are used for forecasting purposes, whether it relates to the epidemic trajectory or final epidemic size. Our findings indicate that careful consideration to the range of epidemic growth profiles arising from infectious disease transmission in specific contact structures is needed during the model development phase. For instance, previous modeling efforts have incorporated mechanisms that indirectly support slower epidemic growth patterns in the context of several infectious diseases including HIV/AIDS [36], Ebola [51-53], foot-and-mouth disease[26, 43, 44], and smallpox [8, 13, 54].

We note that while the generalized-growth approach described here is useful for characterization of empirical outbreaks and may have applications for disease forecasting, this approach does not provide a straightforward equivalent to the reproduction number, $R_0$ -- except for the special case of exponential growth (p=1). Since $R_0$ has become such an important parameter in the modeling and public health community, a useful area for future theoretical work would be to develop a formal match between the generalized-growth model and ultimately $R_0$, or an equivalent parameter, building on the work of [48] [42, 49].

Our approach has applications beyond the analysis of infectious disease outbreak data. For instance, it could be useful to systematically characterize spatial heterogeneity in different contact networks. Moreover, as model complexity grows from the convolution of detailed epidemiological and large-scale contact networks (e.g., [13, 54-56]), the resulting collective epidemic dynamics are increasingly difficult to characterize as a function of individual model components alone. Hence, the range of epidemic growth profiles generated by exploring a realistic area of the disease parameters space could be used to gauge the plausibility of the model. For example, a transmission model designed to study an epidemic of HIV/AIDS or Ebola and unable to reproduce sub-exponential growth dynamics would automatically fail an initial validation phase, even before application to observed epidemic data.

In the context of limited epidemiological data during the early phase of an infectious disease outbreak, simple phenomenological models, which do not make mechanistic assumptions about the transmission process, have proved useful to generate forecasts of the trajectory of an epidemic (e.g., [21, 57-60]). For instance, the logistic growth model provides a statistical description of a single-outbreak trajectory, inspired from population biology. This model relies on two parameters and follows the equation $C'(t) = r\, C(t)\, [1-C(t)/K]$ where $C'(t)$ describes case incidence over time , "r" is the intrinsic growth rate ( per unit of time), and K is the expected final epidemic size (total no. of cases). This model assumes an initial exponential growth phase that saturates as the number of cases accumulate, a pattern that implicitly captures a gradual decrease in the at-risk susceptible population owing to behavior changes (e.g., campaigns to educate the population on how to avoid contracting the disease) and/or control interventions (e.g., increasing rate of diagnosis and isolation of infectious individuals). However, epidemic models that assume initial exponential growth dynamics are incompatible for forecasting outbreaks characterized by sub-exponential growth dynamics particularly when the "deceleration of growth" parameter is substantially lower than 1.0.

Our analysis of infectious disease outbreak datasets has relied on trajectory matching a generalized-growth model from the field of demography [15] to infectious disease epidemic growth patterns. Various sources of error could confound the reported epidemic growth patterns during the early epidemic phase. For instance, it is well recognized that infectious-disease incidence patterns are prone to underreporting owing to multiple factors including the fact that epidemiological surveillance systems tend to capture more severe cases and often miss asymptomatic or mild cases. Nevertheless, as long as underreporting remains more or less the same (as a constant proportion of true infections), it should be possible to estimate the correct parameters.

In summary, we introduced a generalized-growth model to characterize the early rise of infectious disease outbreaks and revealed significant variation in growth profiles across different pathogens and temporal and social contexts. We find evidence for sub-exponential growth patterns in the majority of the 20 outbreaks we studied, representing both established and emerging viral pathogens with a particular focus on Ebola. We hypothesize that sub-exponential growth may result from highly constrained population contact structures, population immunity, or the early onset of behavior changes or control interventions. Our approach could prove useful to systematically characterize spatial heterogeneity arising from infectious disease spread in specific contact structures, assess shifts in epidemic growth patterns, refine mathematical transmission models, and improve disease forecasting.

**Funding**

We acknowledge financial support from the Division of International Epidemiology and Population Studies, The Fogarty International Center, United States National Institutes of Health, funded in part by the Office of Pandemics and Emerging Threats at the United States Department of Health and Human Services. LS acknowledges support from the European Union Horizon 2020 “Marie Curie” senior fellowship.

## Tables

**Table 1**. Summary of epidemic datasets and estimates of r and p derived from fitting the generalized growth model to the early epidemic growth phase.

| **Disease** | **Outbreak** | **Temporal resolution** | **Ascending phase length (number of data points)** | **Growth rate, r (95%CI)** | **Deceleration of growth parameter, p (95% CI)** | **Data Source** |
|---|---|---|---|---|---|---|
| Pandemic influenza | San Francisco (1918) | Days | 19 | 0.3 (0.27, 0.38) | 0.98 (0.91, 1.0) | [18] |
| Smallpox | Khulna, Bangladesh (1972) | Weeks | 9 | 0.1 (0.08, 0.15) | 0.95 (0.85, 1.0) | [61] |
| Plague | Bombay (1905-06) | Weeks | 9 | 0.11 (0.06, 0.22) | 0.86 (0.68, 1.0) | [62] |
| Measles | London (1948) | Weeks | 9 | 1.76 (1.3, 2.32) | 0.51 (0.47, 0.55) | [63] |
| HIV/AIDS | Japan (1985-2012) | Years | 11 | 10.05 (8.02, 12.97) | 0.5 (0.47, 0.54) | [64] |
| AIDS | NYC (1982-2002) | Years | 11 | 19.62 (18.33, 20.0) | 0.57 (0.57, 0.58) | [65] |
| FMD | Uruguay (2001) | Days | 11 | 3.2 (1.74, 5.19) | 0.42 (0.27, 0.58) | [26, 66] |
| Ebola | Uganda (2000) | Weeks | 6 | 0.33 (0.2, 0.51) | 0.68 (0.54, 0.84) | [7, 67] |
| Ebola | Congo (1976) | Days | 20 | 1.43 (0.69, 2.62) | 0.4 (0.22, 0.6) | [68, 69] |
| Ebola | Gueckedou, Guinea (2014) | Weeks | 11 | 0.12 (0.04, 0.3) | 0.71 (0.31, 1.0) | [70] |
| Ebola | Montserrado, Liberia (2014) | Weeks | 10 | 0.27 (0.16, 0.4) | 0.6 (0.46, 0.77) | [70] |
| Ebola | Margibi, Liberia (2014) | Weeks | 8 | 0.11 (0.1, 0.16) | 0.96 (0.80, 1.0) | [70] |
| Ebola | Bomi, Liberia (2014) | Weeks | 8 | 1.2 (0.49, 2.0) | 0.14 (0, 0.37) | [70] |
| Ebola | Grand Bassa, Liberia (2014) | Weeks | 9 | 0.35 (0.1, 0.79) | 0.42 (0.09, 0.83) | [70] |
| Ebola | Western Area Urban, Sierra Leone (2014) | Weeks | 10 | 0.68 (0.39, 1.1) | 0.46 (0.35, 0.59) | [70] |
| Ebola | Western Area Rural, Sierra Leone | Weeks | 10 | 0.26 (0.18, 0.36) | 0.68 (0.59, 0.79) | [70] |
| | (2014) | | | | | |
| Ebola | Bo, Sierra Leone (2014) | Weeks | 10 | 0.08 (0.07, 0.11) | 0.97 (0.86, 1.0) | [70] |
| Ebola | Bombali, Sierra Leone (2014) | Weeks | 8 | 0.09 (0.06, 0.15) | 0.94 (0.74, 1.0) | [70] |
| Ebola | Kenema, Sierra Leone (2014) | Weeks | 8 | 2.38 (1.18, 4.12) | 0.21 (0.06, 0.36) | [70] |
| Ebola | Port Loko, Sierra Leone (2014) | Weeks | 8 | 0.65 (0.37, 1.05) | 0.47 (0.33, 0.61) | [70] |

FMD=foot-and-mouth disease

**Figures**

**Figure 1**. Simulated profiles of epidemic growth (case incidence and relative growth rates) supported by the generalized growth model. The deceleration parameter p is varied while parameter r is fixed at 1.5 per day and C(0)=5.

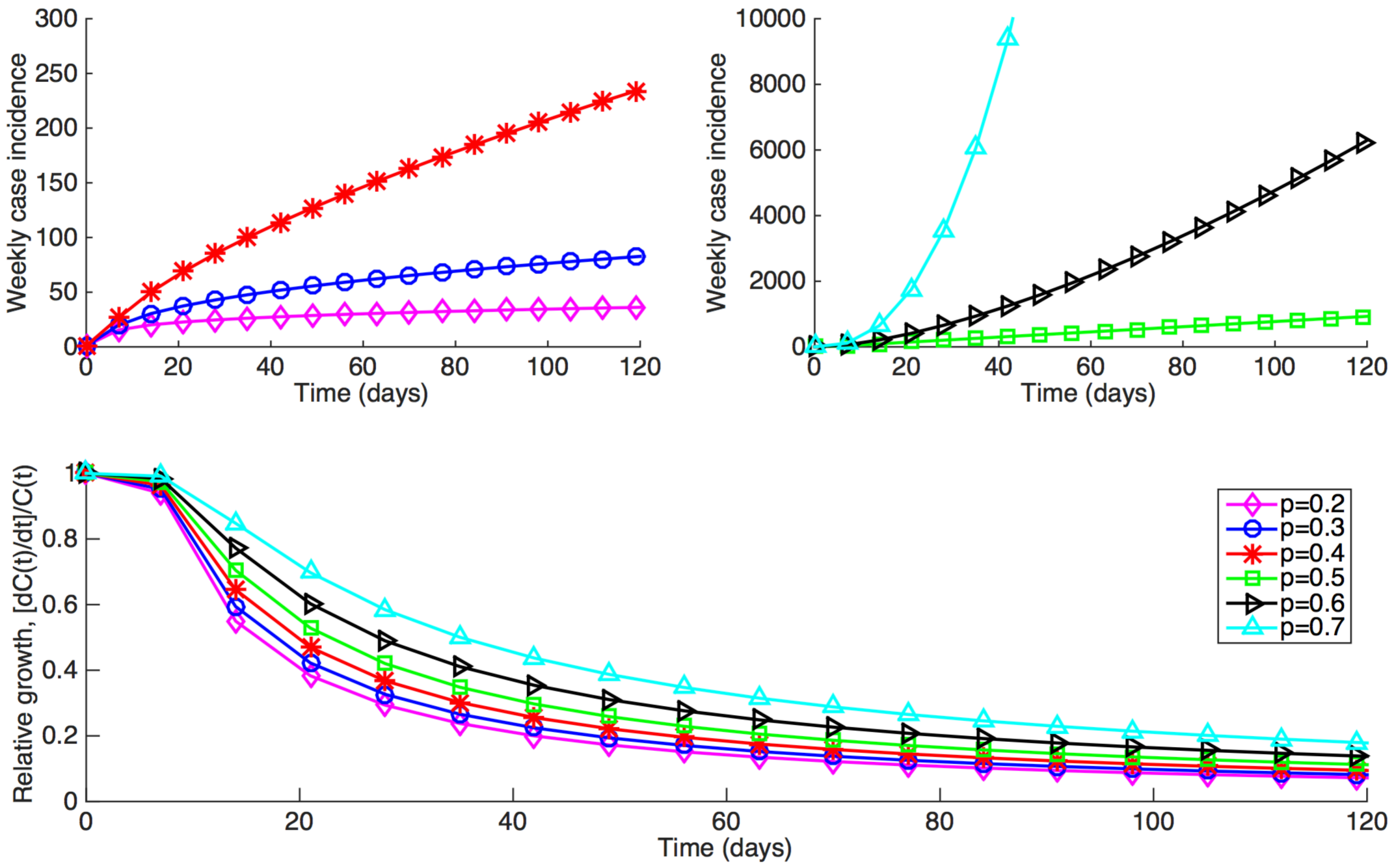

**Figure 2**. Simulations assessing sensitivity of cumulative cases to small variations of the parameter p (0.46-0.62) in the generalized-growth model while fixing parameter r at 1.5 per day and C(0)=5.

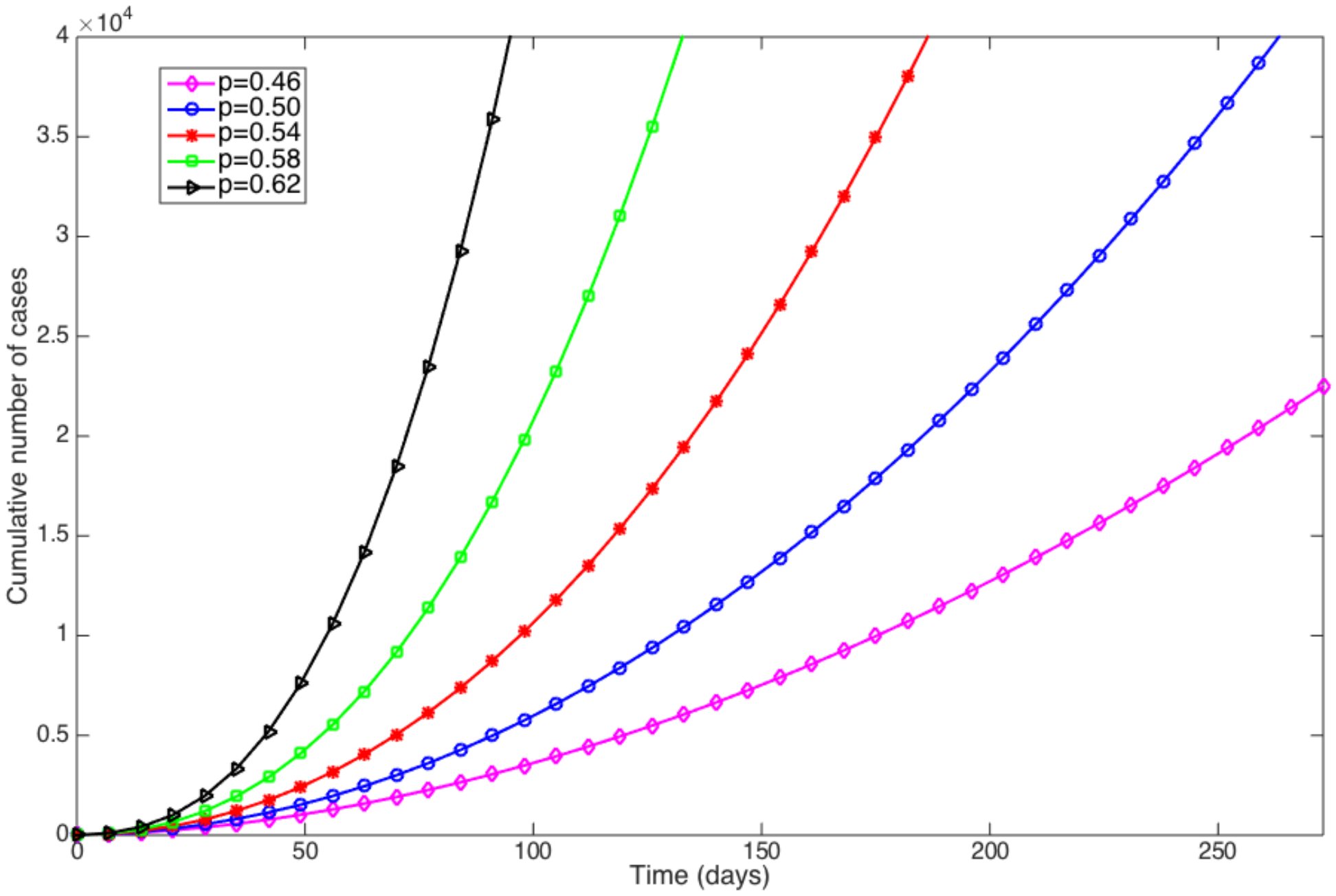

**Figure 3**. Estimates of p and corresponding 95% confidence intervals derived from various infectious disease outbreak datasets of case incidence series by fitting the generalized-growth model to the initial phase of the epidemics as explained in the text. The vertical dashed line separates Ebola and non-Ebola outbreak estimates.

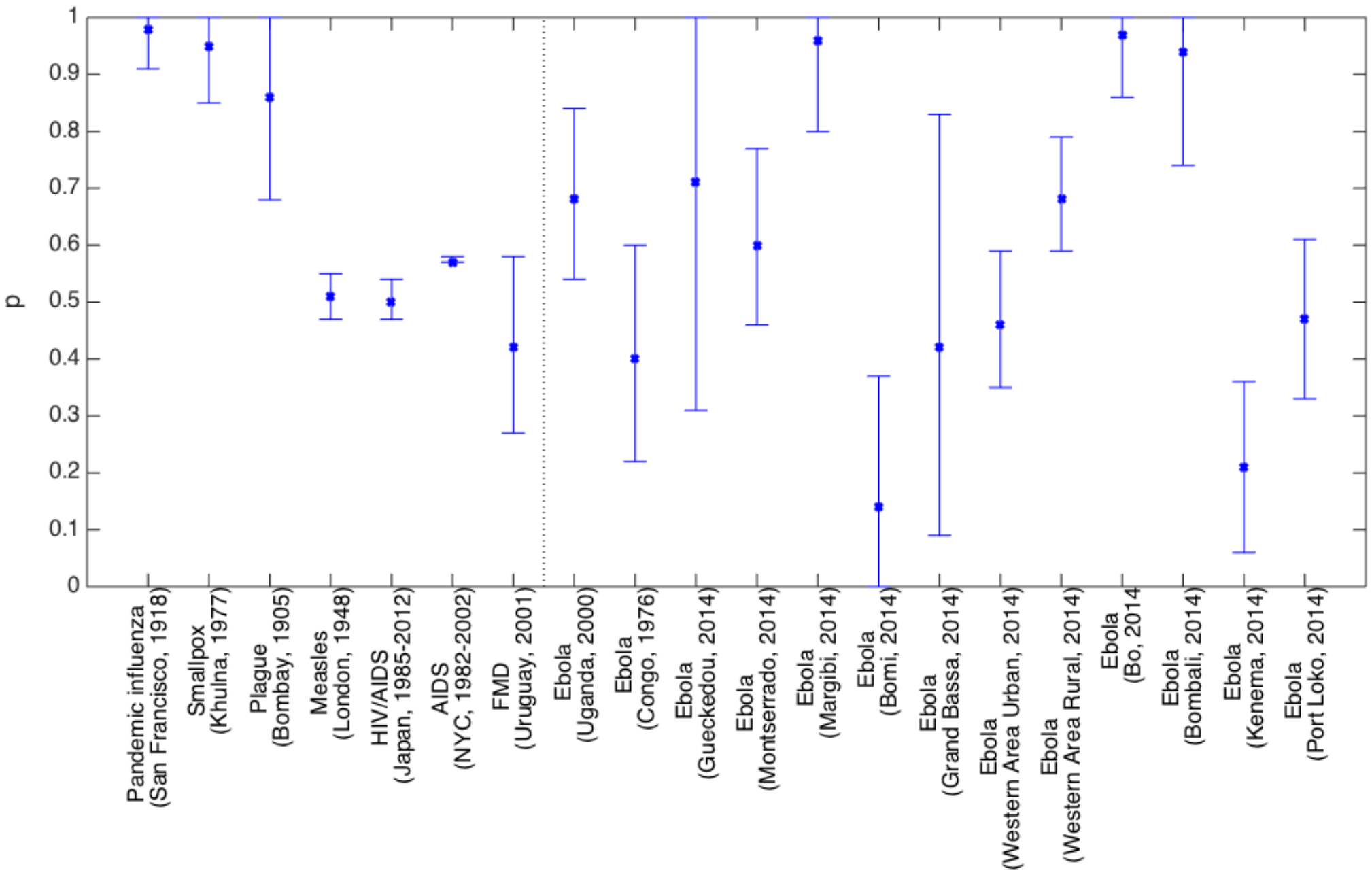

**Figure 4**. **The 1918 influenza pandemic in San Francisco**. Estimates and 95% confidence intervals for parameters r and p obtained by nonlinear least-square fitting the generalized growth model to an increasing amount of case incidence data during the initial epidemic growth phase are shown in the first two panels. The statistical comparisons of the generalized-growth model fit to the simpler exponential growth model where p=1 (gray shaded periods indicate periods where the generalized-growth model provides a better fit compared to the exponential growth model) are also shown in the upper right panel. Representative fits of the generalized-growth model to various epidemic growth phases are displayed in the bottom panels.

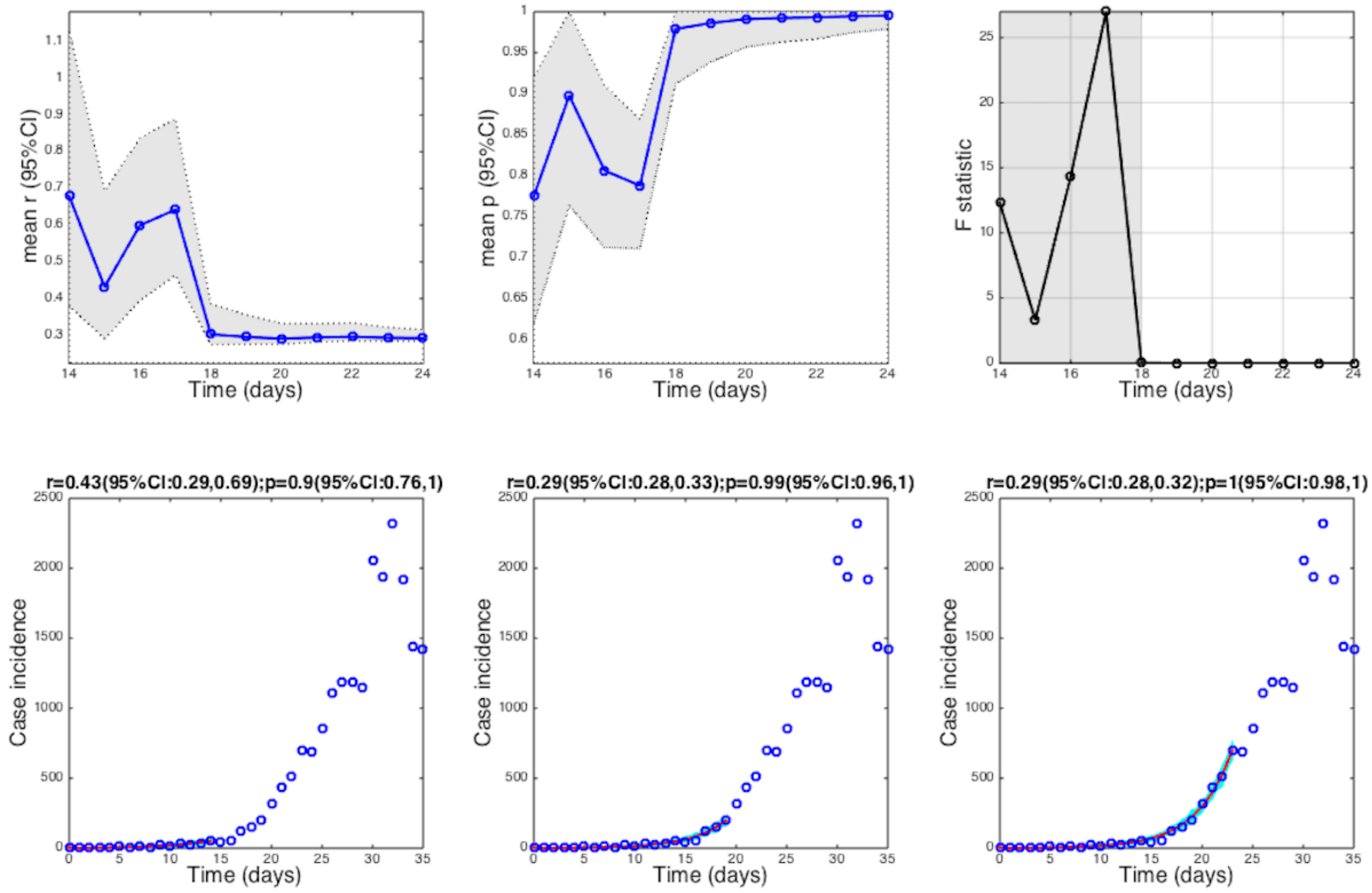

**Figure 5. Smallpox epidemic in Khulna municipality, Bangladesh in 1972.** Estimates and 95% confidence intervals for parameters r and p obtained by nonlinear least-square fitting the generalized growth model to an increasing amount of case incidence data during the initial epidemic growth phase are shown in the first two panels. The statistical comparisons of the generalized-growth model fit to the simpler exponential growth model where p=1 (gray shaded periods indicate periods where the generalized-growth model provides a better fit compared to the exponential growth model) are also shown in the upper right panel. Representative fits of the generalized-growth model to various epidemic growth phases are displayed in the bottom panels.

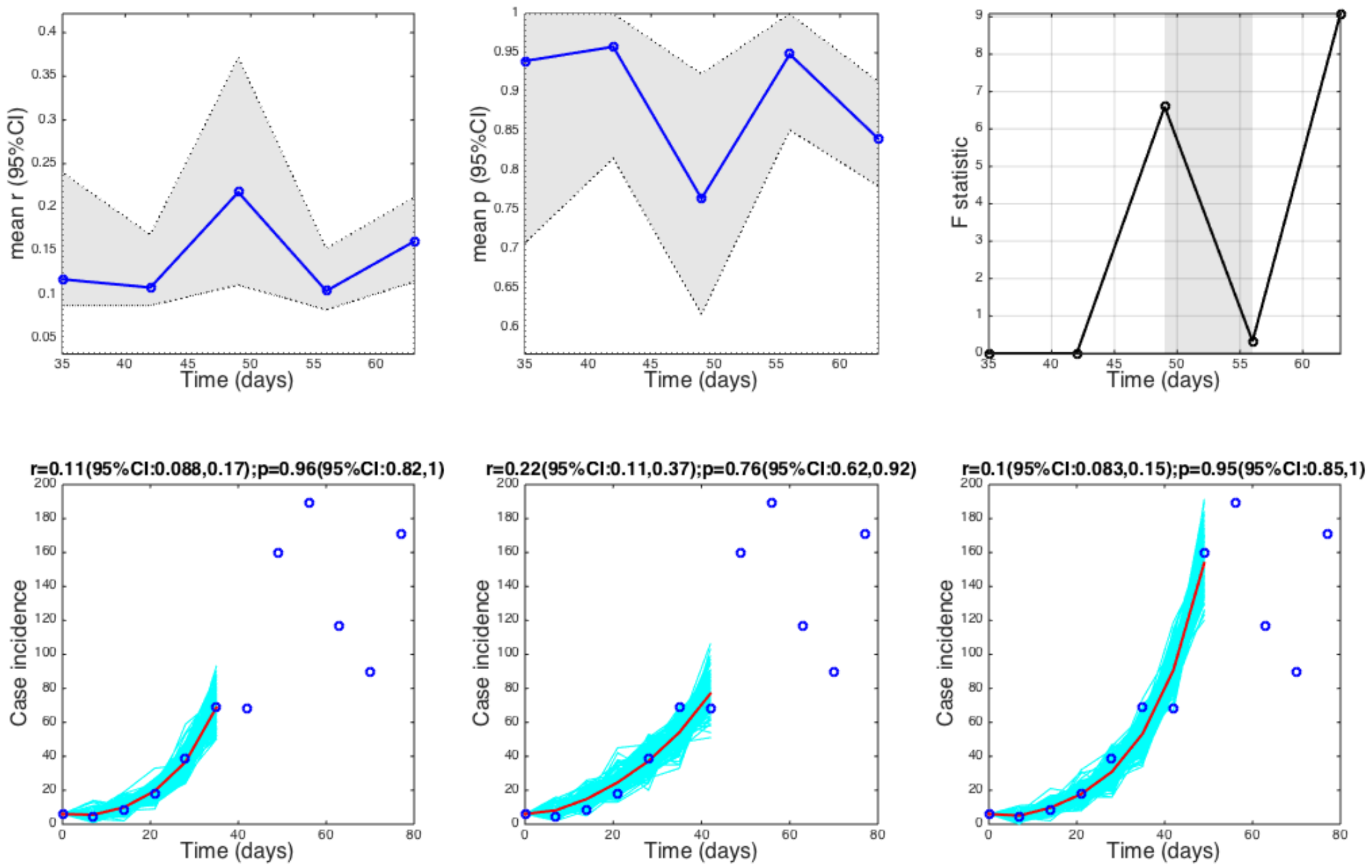

**Figure 6. The 2001 foot-and-mouth disease epidemic in Uruguay.** Estimates and 95% confidence intervals for parameters r and p obtained by nonlinear least-square fitting the generalized growth model to an increasing amount of case incidence data during the initial epidemic growth phase are shown in the first two panels. The statistical comparisons of the generalized-growth model fit to the simpler exponential growth model where p=1 (gray shaded periods indicate periods where the generalized-growth model provides a better fit compared to the exponential growth model) are also shown in the upper right panel. Representative fits of the generalized-growth model to various epidemic growth phases are displayed in the bottom panels.

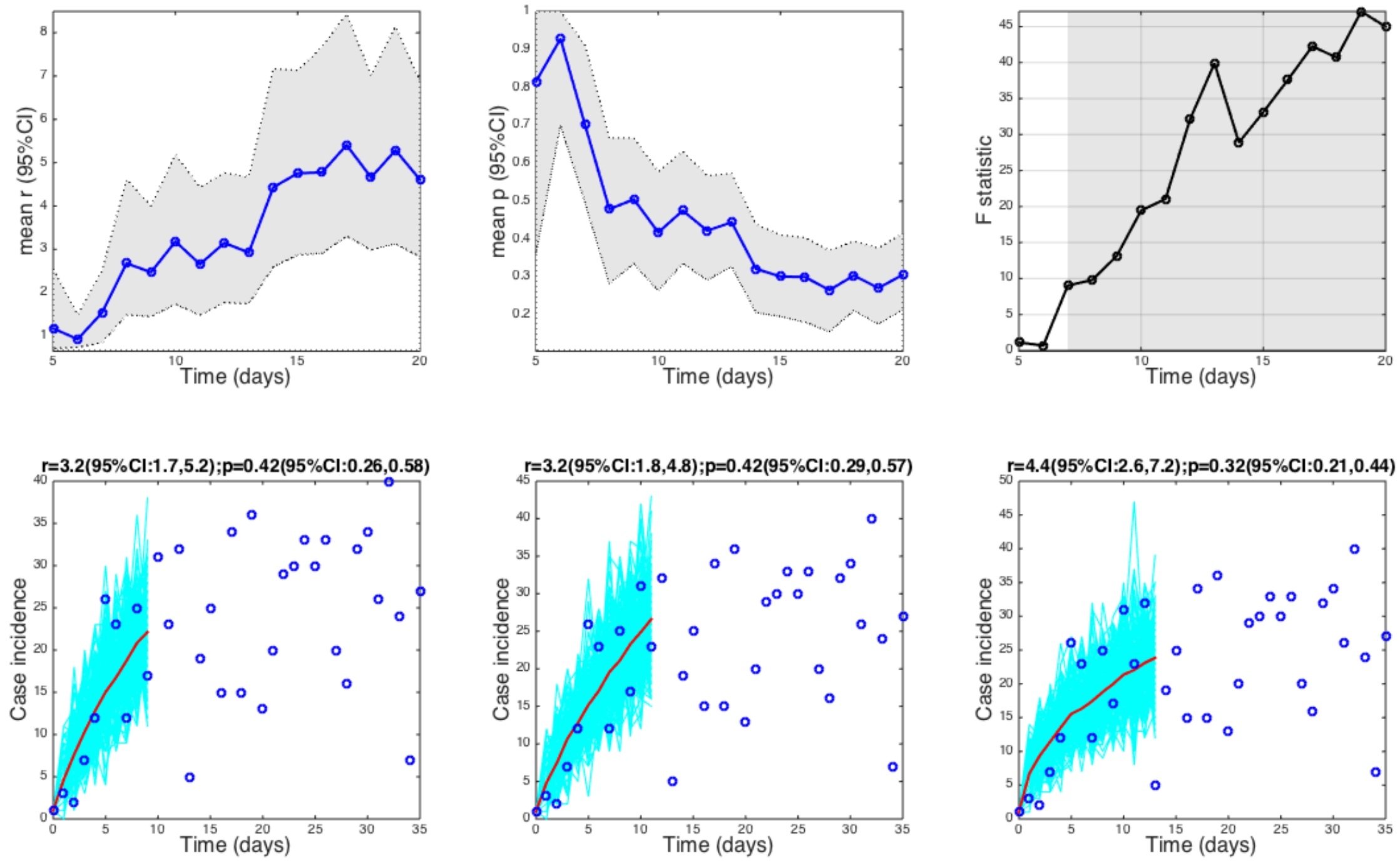

**Figure 7. The HIV/AIDS epidemic in Japan (1985-2012).** Estimates and 95% confidence intervals for parameters r and p obtained by nonlinear least-square fitting the generalized growth model to an increasing amount of case incidence data during the initial epidemic growth phase are shown in the first two panels. The statistical comparisons of the generalized-growth model fit to the simpler exponential growth model where p=1 (gray shaded periods indicate periods where the generalized-growth model provides a better fit compared to the exponential growth model) are also shown in the upper right panel. Representative fits of the generalized-growth model to various epidemic growth phases are displayed in the bottom panels.

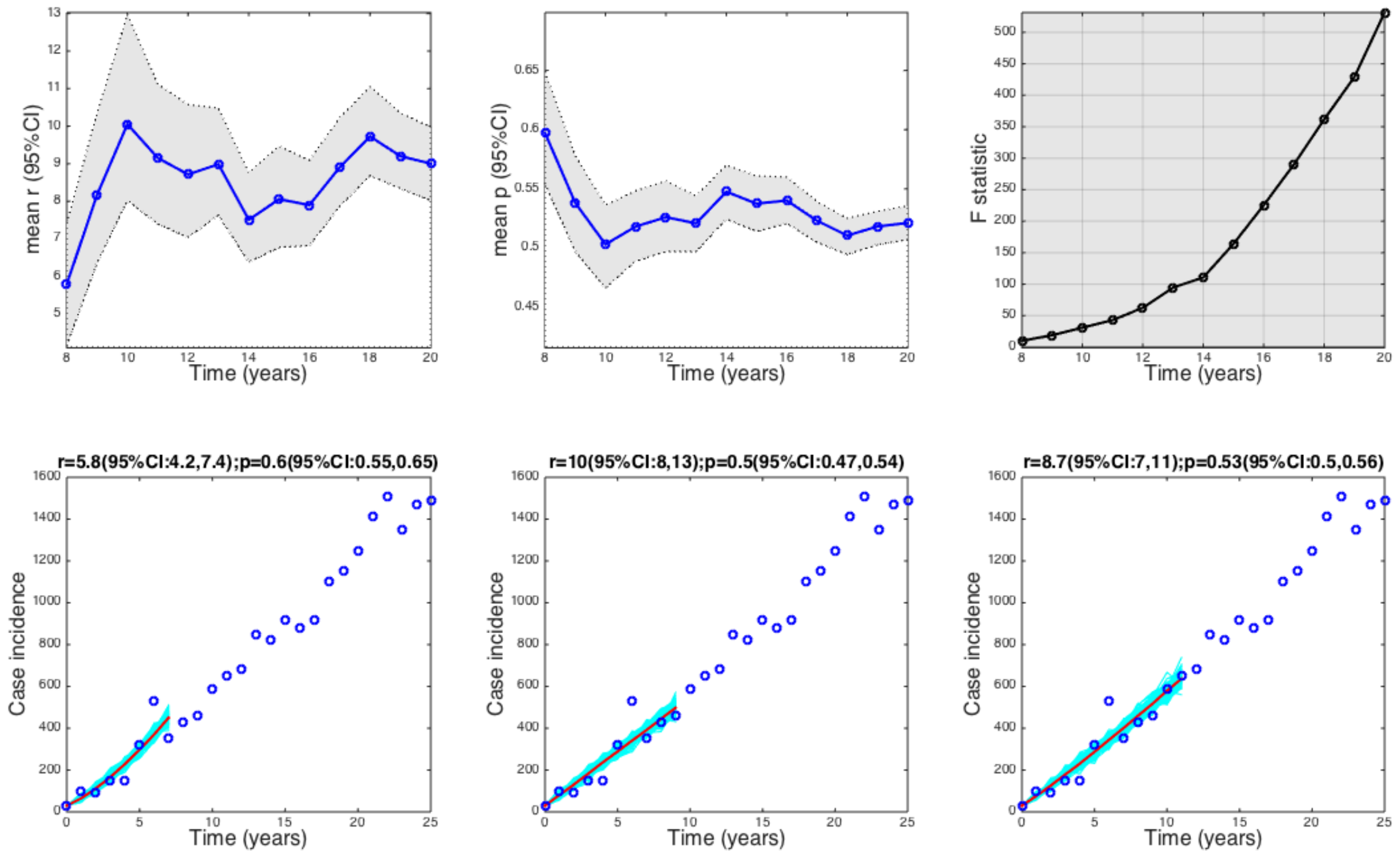

**Figure 8. The 2014-15 Ebola epidemic in Montserrado, Liberia.** Estimates and 95% confidence intervals for parameters r and p obtained by nonlinear least-square fitting the generalized growth model to an increasing amount of case incidence data during the initial epidemic growth phase are shown in the first two panels. The statistical comparisons of the generalized-growth model fit to the simpler exponential growth model where p=1 (gray shaded periods indicate periods where the generalized-growth model provides a better fit compared to the exponential growth model) are also shown in the upper right panel. Representative fits of the generalized-growth model to various epidemic growth phases are displayed in the bottom panels.

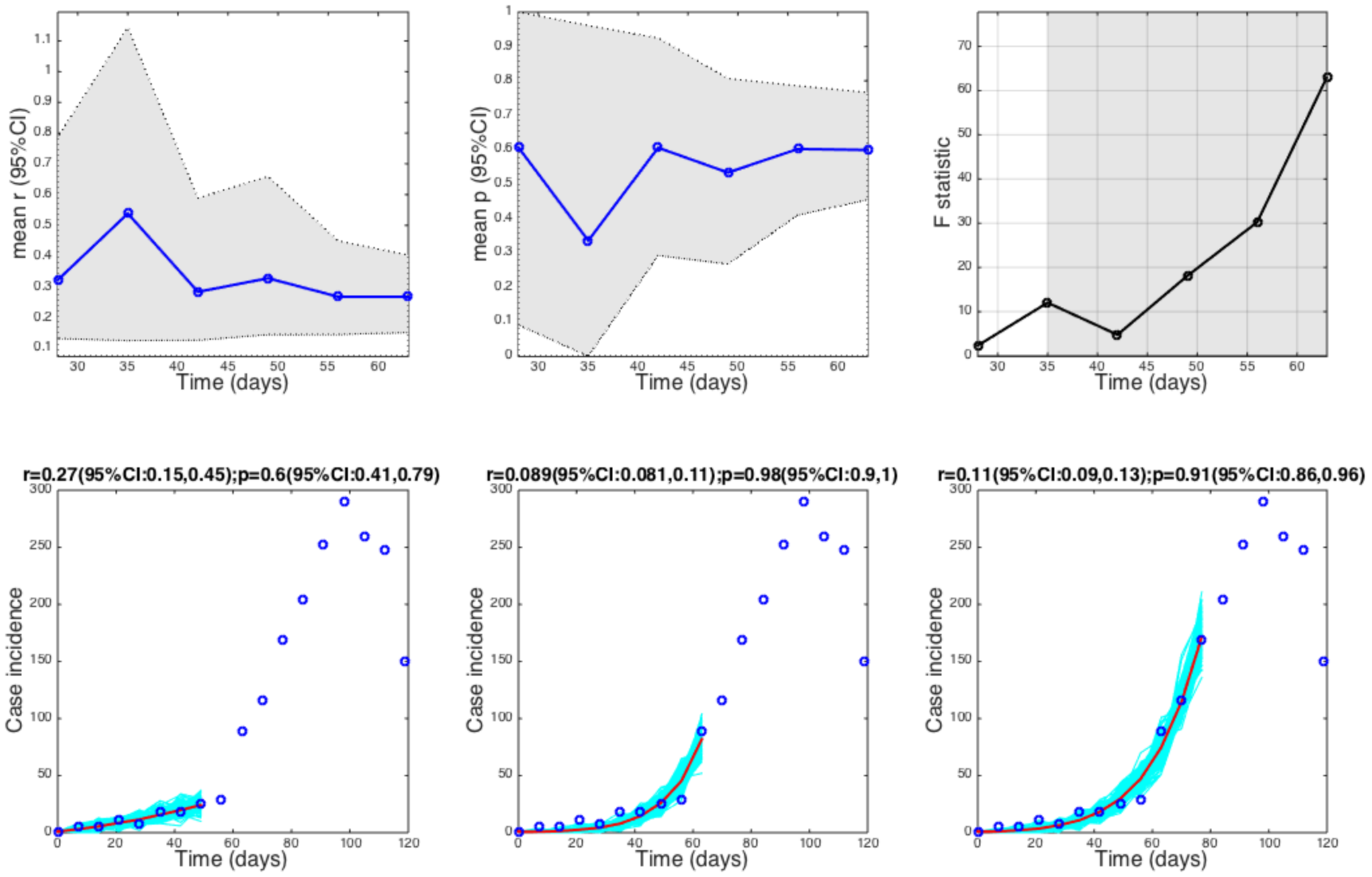

**Figure 9. The 2014-15 Ebola epidemic in Western Area Urban, Sierra Leone.** Estimates and 95% confidence intervals for parameters r and p obtained by nonlinear least-square fitting the generalized growth model to an increasing amount of case incidence data during the initial epidemic growth phase are shown in the first two panels. The statistical comparisons of the generalized-growth model fit to the simpler exponential growth model where p=1 (gray shaded periods indicate periods where the generalized-growth model provides a better fit compared to the exponential growth model) are also shown in the upper right panel. Representative fits of the generalized-growth model to various epidemic growth phases are displayed in the bottom panels.

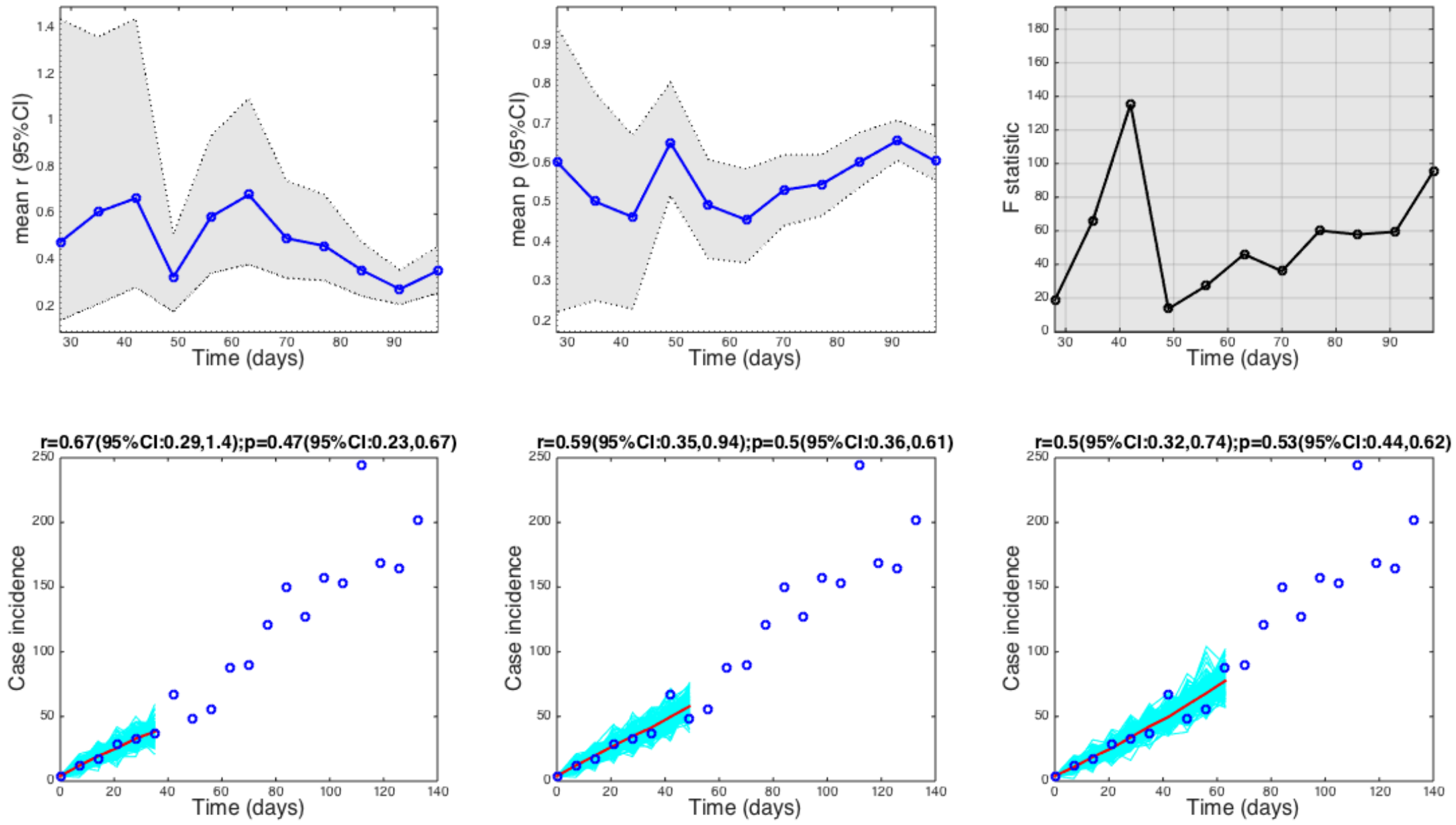

**Supplementary Figure S1.** Estimates of “r” and “p” from simulated data of the generalized-growth model. Results indicate that parameter uncertainty as measured by the width of the 95% confidence interval of the parameter estimates is reduced by larger case numbers, so that it typically remains higher the smaller the “deceleration of growth” parameter p. As expected, parameter uncertainty is reduced as more cases are reported over time during the initial epidemic growth phase. Simulations of the generalized-growth model were used for values of p at 0.3, 0.4, 0.5, and 0.6 while keeping value of r fixed at 1.5.

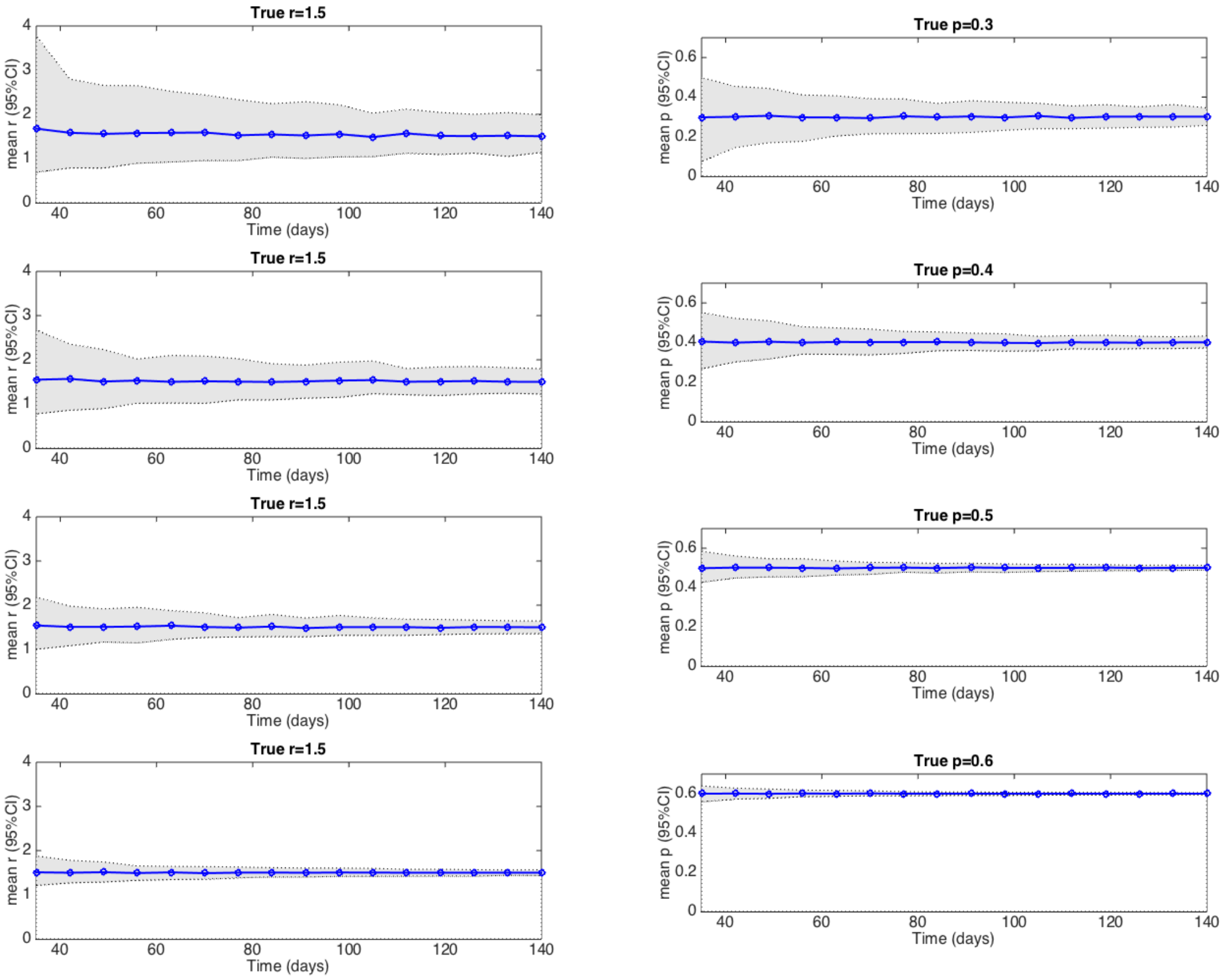